\documentclass[10pt, twocolumn, notitlepage, superscriptaddress, prb, longbibliography]{revtex4-2}

\usepackage{here}
\usepackage{amsmath}         
\usepackage{graphicx}
\usepackage{braket}         
 \usepackage{lettrine}
 \usepackage{color}
 \usepackage{times}
\usepackage{hhline}
\usepackage{tabularx}
\usepackage{calc}
\usepackage{soul}

\begin{document}

\title{Orbital Rashba effect in surface oxidized Cu film}    

\author{Dongwook Go}
\email{d.go@fz-juelich.de}
\affiliation{Peter Gr\"unberg Institut and Institute for Advanced Simulation, Forschungszentrum J\"ulich and JARA, 52425 J\"ulich, Germany \looseness=-1}
\affiliation{Institute of Physics, Johannes Gutenberg University Mainz, 55099 Mainz, Germany}

\author{Daegeun Jo}
\affiliation{Department of Physics, Pohang University of Science and Technology, Pohang 37673, Korea \looseness=-1}

\author{Tenghua Gao}
\affiliation{Department of Applied Physics and Physico-Informatics, Keio University, Yokohama 223-8522, Japan}
\affiliation{Keio Institute of Pure and Applied Sciences (KiPAS), Keio University, Yokohama 223-8522, Japan}

\author{Kazuya Ando}
\affiliation{Department of Applied Physics and Physico-Informatics, Keio University, Yokohama 223-8522, Japan}
\affiliation{Keio Institute of Pure and Applied Sciences (KiPAS), Keio University, Yokohama 223-8522, Japan}
\affiliation{Center for Spintronics Research Network (CSRN), Keio University, Yokohama 223-8522, Japan}



\author{Stefan Bl\"ugel}
\affiliation{Peter Gr\"unberg Institut and Institute for Advanced Simulation, Forschungszentrum J\"ulich and JARA, 52425 J\"ulich, Germany \looseness=-1}

\author{Hyun-Woo Lee}
\affiliation{Department of Physics, Pohang University of Science and Technology, Pohang 37673, Korea \looseness=-1}

\author{Yuriy Mokrousov}
\affiliation{Peter Gr\"unberg Institut and Institute for Advanced Simulation, Forschungszentrum J\"ulich and JARA, 52425 J\"ulich, Germany \looseness=-1}
\affiliation{Institute of Physics, Johannes Gutenberg University Mainz, 55099 Mainz, Germany}

\begin{abstract}
Recent experimental observation of unexpectedly large current-induced spin-orbit torque in surface oxidized Cu on top of a ferromagnet suggested a possible role of the orbital Rashba effect (ORE). With this motivation, we investigate the ORE from first principles by considering an oxygen monolayer on top of a Cu(111) film. We show that surface oxidization of Cu film leads to gigantic enhancement of the ORE for states near the Fermi surface. The resulting chiral orbital texture in the momentum space is exceptionally strong, reaching $\sim 0.5\hbar$ in  magnitude. We find that resonant hybridization between O $p$-states  and Cu $d$-states is responsible for the emergence of the ORE. We demonstrate that application of an external electric field generates huge orbital Hall current, which is an order of magnitude larger than the spin Hall current found in heavy metals. This implies that ``orbital torque" mechanism may be significant in surface oxidized Cu/ferromagnet structures. It also encourages experimental verification of the orbital texture in surface oxidized Cu through optical measurements such as angle-resolved photoemission spectroscopy.
\end{abstract}

\date{\today}                 
\maketitle	      

The Rashba effect is one of the most important manifestations of spin-orbit coupling (SOC) at surfaces of solids \cite{Rashba1984}. It leads to spin splitting of the electronic band structure, where the direction of the spin is correlated with crystal momentum $\mathbf{k}$. Such spin-momentum locking of the Rashba states attracted huge interest by the  spintronics community owing to its potential use in electrical control of spin \cite{Manchon2015, Koo2020}. By now, various experiments demonstrated the inter-conversion between a charge current and a spin current mediated by the Rashba states \cite{Miron2011, Sanchez2013, Fan2014, Emori2016, Tsai2018, Puebla2019, Haku2020}. 
As for the Rashba effect, it has been revealed that orbital angular momentum (OAM) plays a significant role, which led to the notion of the orbital Rashba effect (ORE) \cite{Park2011, Park2012, Park2013, Go2017, Oh2017, Sunko2017, Unzelmann2020}. The ORE refers to OAM-dependent energy splitting that generates chiral OAM texture in $\mathbf{k}$ space. Despite its similarity to the SRE, the mechanism of the ORE is independent of SOC, with the inversion symmetry breaking alone sufficient for its emergence. As the SOC is taken into account, the spin moment couples to the chiral OAM texture formed by the ORE, leading to coexistence of the SRE and ORE. In this regard, the ORE can be considered more fundamental than the SRE. The ORE and resulting OAM texture have been studied in various systems including not only prototypical noble metal surfaces \cite{Kim2012} but also metal surface alloys \cite{Go2017, Unzelmann2020}, topological surface state \cite{Park2012b}, bulk ferroelectrics \cite{Ponet2018}, oxide surface \cite{Sunko2017}, and two-dimensional materials \cite{Canonico2020}.

Meanwhile, An and co-workers report an observation of an unusually large current-induced spin-orbit torque (SOT) in a surface oxidized Cu/ferromagnet structure \cite{An2016}, where electric current flowing parallel to the interface plane induces dynamics of the ferromagnetic moment. Considering that SOT is mostly observed in materials containing heavy elements with large SOC, this result is unconventional. 
This gives a clue that the OAM may play a role because electric generation of the orbital current does not require any SOC \cite{Bernevig2005, Go2018, Canonico2020, Bhowal2020}. At the same time, Ref.~\cite{Gao2018} strongly supports that an interfacial mechanism is responsible for the SOT when the oxidized Cu is electrically insulating. One of the proposed mechanisms suggests a crucial role of the ORE in surface oxidized Cu, by which "orbital current" is generated and injected into the adjacent ferromagnet \cite{Kim2020, Tazaki2020, Ding2020}. However, another mechanism based on spin-vorticity coupling has also been proposed \cite{Okano2019}. To our best knowledge, however, neither direct evidence of the ORE nor theoretical investigation in surface oxidized Cu has been given. Since torque generation mechanism based on the orbital current involves more complicated conversion processes \cite{Go2020}, it is necessary to understand orbital-related properties of surface oxidized Cu itself without additional ferromagnetic layer.

\begin{figure}[t!]
\includegraphics[angle=0, width=0.45\textwidth]{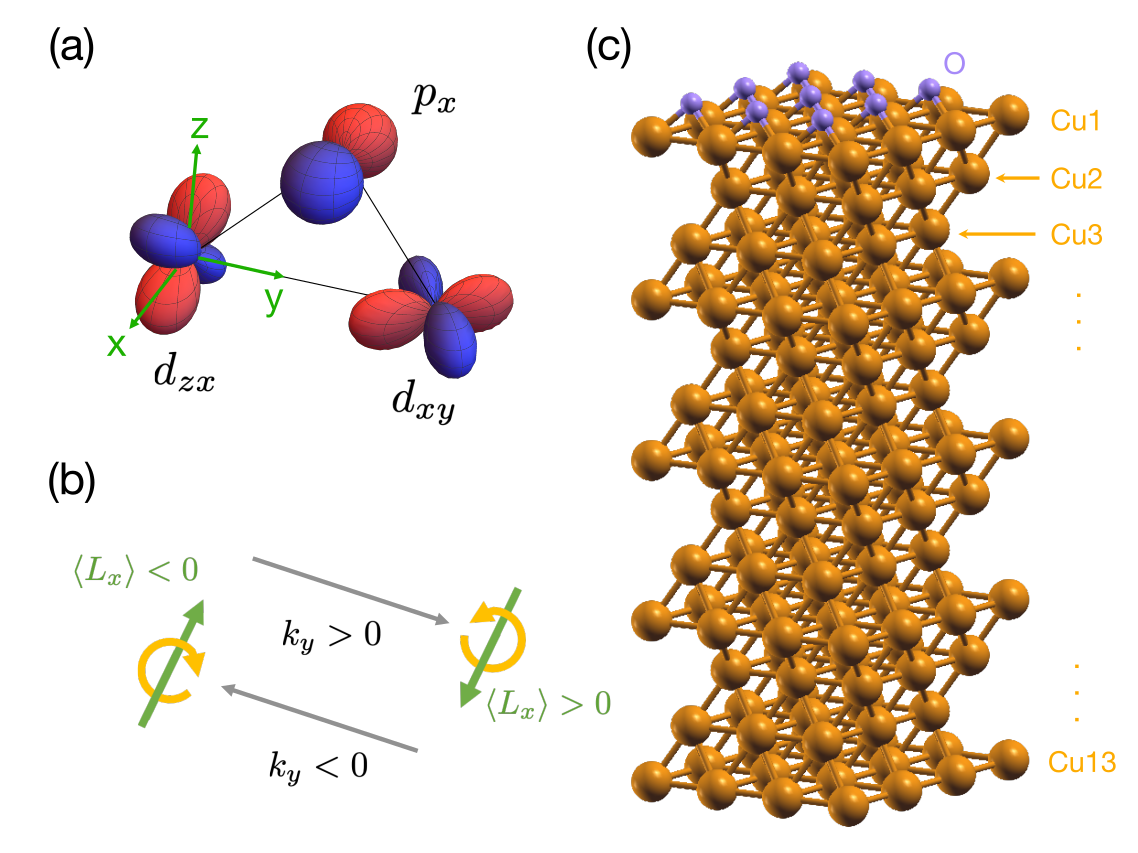}
\caption{
\label{fig:structure_mechanism}
(a) Crystal structure of surface oxidized Cu. Purple and orange spheres represent O and Cu atoms, respectively. (b) Hybridization between an O $p$-orbital and Cu $d$-orbitals mediates the formation of a superposition of the type  $d_{zx}\pm i d_{xy}$ , which carries finite OAM. (c) As a result, electrons moving in opposite directions have an opposite sign of OAM expectation value, which constitutes the ORE.
}
\end{figure}

In this Rapid Communication, we answer to questions whether the ORE is induced by surface oxidization of a Cu film and whether it can generate sufficiently large orbital current as an external electric field is applied. From first principles calculation based on the density functional theory, we investigate an oxygen monolayer on top of a Cu(111) film as a model for surface oxidized Cu [Fig.~\ref{fig:structure_mechanism}(a)]. We show that surface oxidization of the Cu(111) film indeed induces gigantic ORE compared to bare Cu(111) surface. The resulting chiral OAM in the $\mathbf{k}$ space is exceptionally large reaching as much as $\sim 0.5\hbar$ in its magnitude. As a mechanism behind the gigantic ORE, we demonstrate that resonant hybridization between O $p$ orbitals and Cu $d$ orbitals is responsible. We further find that an external electric field generates not only OAM accumulation but also orbital Hall current, whose magnitude surpasses spin Hall current observed in heavy elements. This implies that "orbital torque" mechanism may be significant in surface oxidized Cu/ferromagnet structures. It also encourages experimental verification of the orbital texture in surface oxidized Cu through optical measurements such as angle-resolved photoemission spectroscopy.

An experiment revealed strong correlation between oxygen-induced orbital hybridization and SOT \cite{Kageyama2019}. Here, we show that $pd$ orbital hybridization bewteen O and Cu atoms leads to emergence of the ORE. To illustrate this point, we consider a model with a minimal  orbital basis: $p_y$ orbital from O and $d_{zx}$ and $d_{xy}$ orbitals from Cu [Fig.~\ref{fig:structure_mechanism}(b)]. We further assume that the atoms lie within the $yz$ plane and consider only $y$ directional motion of the electron for simplicity. Note that the superposition of $d_{zx}$ and $d_{xy}$ orbitals, such as $d_{zx}\pm id_{xy}$, may carry finite OAM along the $x$ direction. Taking only nearest neighbor hoppings into account, the Hamiltonian near $k_y=0$ can be expressed as
\begin{eqnarray}
\mathcal{H} (k_y)
=
\begin{pmatrix}
E_d & 0 & \alpha \\
0 & E_d &  i \beta k_y \\
\alpha & -i\beta k_y  & E_p
\end{pmatrix},
\end{eqnarray}
where basis states are Bloch states with $d_{zx}$, $d_{xy}$, and $p_y$ orbital character. Here, $E_{d(p)}$ is the onsite energy of $d(p)$ orbital, and $\alpha(\beta)$ parameterizes the nearest hopping amplitudes between $p_y$ and $d_{zx}(d_{xy})$ orbitals, which are real numbers. By downfolding the $pd$ hybridization (terms proportional to $\alpha$ and $\beta$), we can obtain an effective model in the $d$ orbital space. Assuming that the energy gap between $p$ and $d$ orbitals is larger than $\alpha$ and $\beta k_y$, a $k_y$-linear additional term appearing in the $d$ orbital space is found as 
\begin{eqnarray}
\mathcal{H}'_{d} (k_y)
=
-\frac{\alpha\beta k_y}{\hbar(E_d - E_p)}
L_x,
\end{eqnarray}
where 
\begin{eqnarray}
L_x = \hbar 
\begin{pmatrix}
0 & i \\ 
-i & 0
\end{pmatrix}
\end{eqnarray}
is the $x$ component of the OAM operator in the $d$ orbital sector. As a result, electrons moving in opposite directions have an opposite sign of OAM expectation value, as illustrated in Fig.~\ref{fig:structure_mechanism}(c), which is nothing else but the ORE. One may consider more realistic models, but the conclusion remains same: the $pd$ hybridization mediates an interaction among $d$ states, which induces chiral OAM texture in $\mathbf{k}$-space. Therefore, we can expect that as Cu film becomes oxidized the hybridization between O $p$ orbital and Cu $d$ orbitals would induce the ORE. 

\begin{figure}[t!]
\includegraphics[angle=0, width=0.45\textwidth]{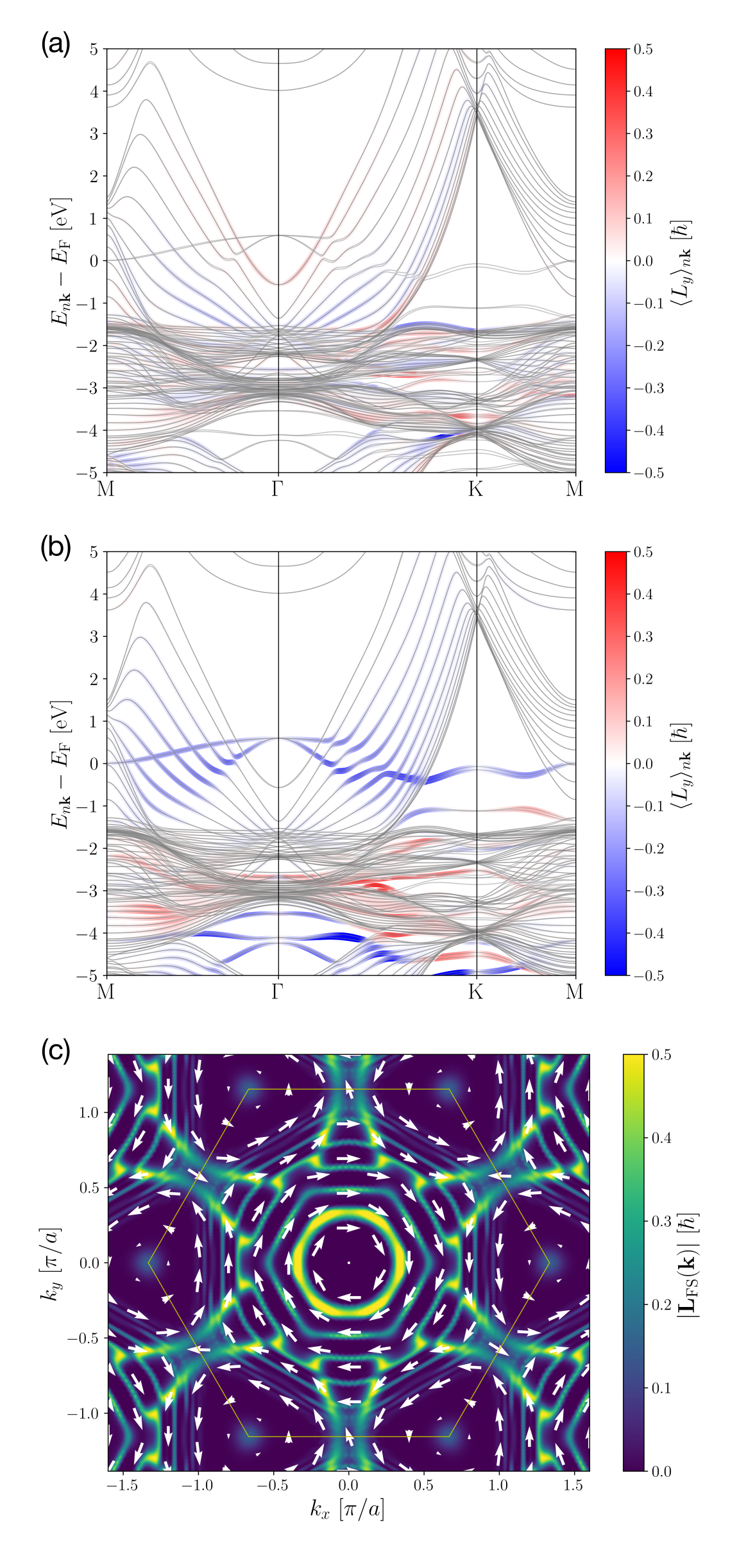}
\caption{
\label{fig:orbital_texture}
(a,b) Electronic band structure of the surface oxidized Cu (grey lines). The color in (a) and (b) indicates the OAM expectation value of each state projected onto the bottom (Cu13) and top (Cu1) Cu layer, respectively. (c) OAM texture at the Fermi surface. The arrows indicates the direction of OAM expectation value and the color represents its absolute value.
}
\end{figure}

The atomic structure of surface oxidized Cu is very complicated in general as it tends to form an amorphous structure \cite{Gattinoni2015}, which is challenging to treat from first principles calculations. A recent experiment indicates that a natural oxdization leads to a duplex type structure CuO/Cu$_2$O/Cu \cite{An2016}. However, another experiment performed on AlO$_x$/Cu structure implies that formation of  Cu-oxide is not crucial for the properties of oxidized Cu. This suggests that oxygen itself is a more important element than the Cu oxide. Such ``free" oxygen can efficiently hybridize with Cu, which is also implied in the model in Fig.~\ref{fig:structure_mechanism}(b). On the other hand, inversion symmetry breaking at the interface between Cu-oxide/Cu is expected to be weak due to lack of the free oxygen. In fact, an experiment shows that this structure is far from generating substantial SOT \cite{An2016}.

With these considerations, we model the surface oxidized Cu as an oxygen monolyaer on top of Cu(111) film as shown in Fig. \ref{fig:structure_mechanism}(a).
Figures~\ref{fig:orbital_texture}(a) and \ref{fig:orbital_texture}(b) show the electronic band structure, where the color indicates the expectation value of $y$-component of the OAM for each state projected onto the bottom Cu layer (Cu13) and the tom Cu layer (Cu1), respectively. In Fig.~\ref{fig:orbital_texture}(a), the bottom surface state, which is located around $-0.4\ \mathrm{eV}$ below the Fermi energy, shows an orbital texture with a magnitude of $\sim 0.1 \hbar$. Spin splitting is almost negligible due to small SOC of Cu, leading to almost doubly degenerate bands. The ORE in a bare Cu(111) surface has been previously investigated through circular dichroism measurement \cite{Kim2012}. On the top surface, however, we find additional states which emerge from the hybridization with the oxygen layer. These states exhibit gigantic OAM expectation value reaching as much as $\sim 0.5\hbar$, which is much larger than for any other known Rashba system \cite{Go2017, Sunko2017}. It is important to notice that the states responsible for the pronounced OAM texture do not exist in bare Cu(111) surface \cite{Ishida2014}.

Since OAM of the states near the Fermi surface is the most relevant for electronic transport, we evaluate the OAM texture at the Fermi surface. We calculate the Fermi surface OAM expectation value by 
\begin{eqnarray}
\mathbf{L}_\mathrm{FS}(\mathbf{k})
&=&
-4 k_\mathrm{B} T \sum_n f'_{n\mathbf{k}}
\left\langle \mathbf{L} \right\rangle_{n\mathbf{k}}
\nonumber
\\
&=&
\sum_n \frac{2
\left\langle \mathbf{L} \right\rangle_{n\mathbf{k}}
}{1+\cosh [(E_\mathrm{F}-E_{n\mathbf{k}})/k_\mathrm{B}T]}
,
\end{eqnarray}
where $f'_{n\mathbf{k}}$ is energy derivative of the Fermi-Dirac distribution function, $k_\mathrm{B}$ is the Boltzmann constant, and $T$ is the temperature so that $k_BT = 25$\,meV. Note that this definition makes the maximum value of $\left\langle \mathbf{L} \right\rangle_{n\mathbf{k}}$ coincides with the maximum value of $\mathbf{L}_\mathrm{FS}(\mathbf{k})$ at the Fermi energy.
Figure~\ref{fig:orbital_texture}(c) shows the distribution of $\mathbf{L}_\mathrm{FS}(\mathbf{k})$ in $\mathbf{k}$-space, where the eigenstates are projected onto Cu1 layer for the evaluation of $\left\langle \mathbf{L} \right\rangle_{n\mathbf{k}}$. The color represents the magnitude and the arrows indicate the direction of the OAM. We observe that  not only does the OAM pattern show pronounced values near the Fermi surface,  but that it also exhibits a clear chirality, which is clockwise around the $\Gamma$-point. These results confirm that the ORE becomes pronounced by surface oxidation of Cu(111).

\begin{figure}[t!]
\includegraphics[angle=0, width=0.45\textwidth]{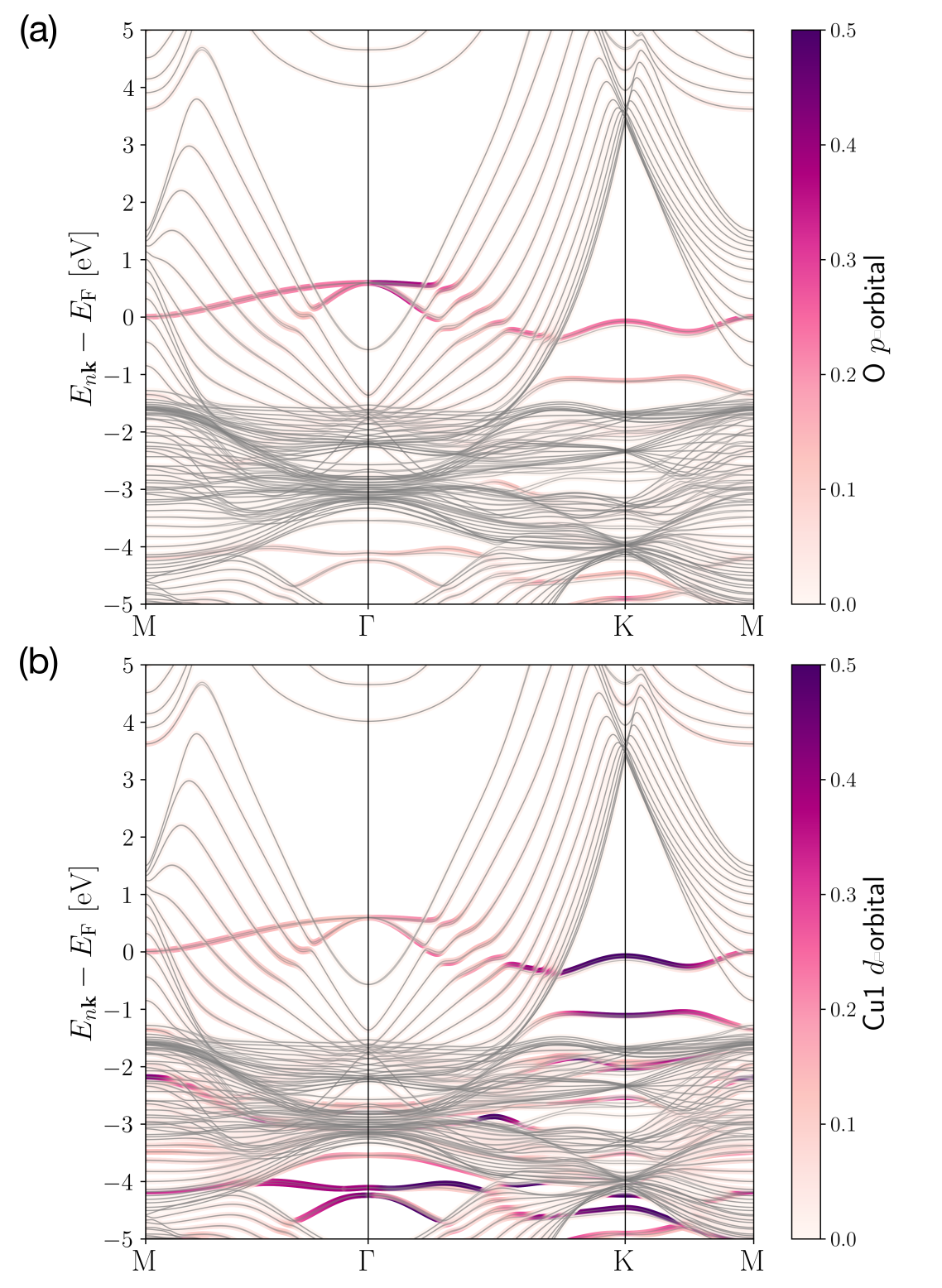}
\caption{
\label{fig:character_analysis}
Orbital character analysis. On top of the energy bands (grey lines), the weight of (a) O $p$ and (b) Cu1 $d$ orbitals is shown with the color code.
}
\end{figure}

To confirm whether the ORE is indeed triggered by the $pd$ hybridization as sketched in Fig.~\ref{fig:structure_mechanism}(b), we analyze the orbital character of the electronic bands. Figures~\ref{fig:character_analysis}(a) and \ref{fig:character_analysis}(b) show the weights of O $p$ and Cu1 $d$ orbitals, respectively, for each state. By comparing the weights with Fig.~\ref{fig:orbital_texture}(b), we find that the states responsible for the pronounced OAM are the states with dominant O $p$ and Cu1 $d$ orbital characters. Therefore, this confirms our expectation that the $pd$ hybridization induces the ORE in surface oxidized Cu. It is remarkable that Cu1 $d$ character is promoted near the Fermi energy as a result of the $pd$ hybridization. In bulk Cu, the states with a sizeable $d$ character are found at the energies which are at least $1.5\ \mathrm{eV}$ lower than the Fermi level. This implies that the magnitude of the $pd$ hybridization is strong enough to change the energy of the $d$-state significantly. This result explains variation of the SOT and magnetic anisotropy with respect to change of the oxygen concentration at the interface \cite{Kageyama2019}.

\begin{figure}[t!]
\includegraphics[angle=0, width=0.45\textwidth]{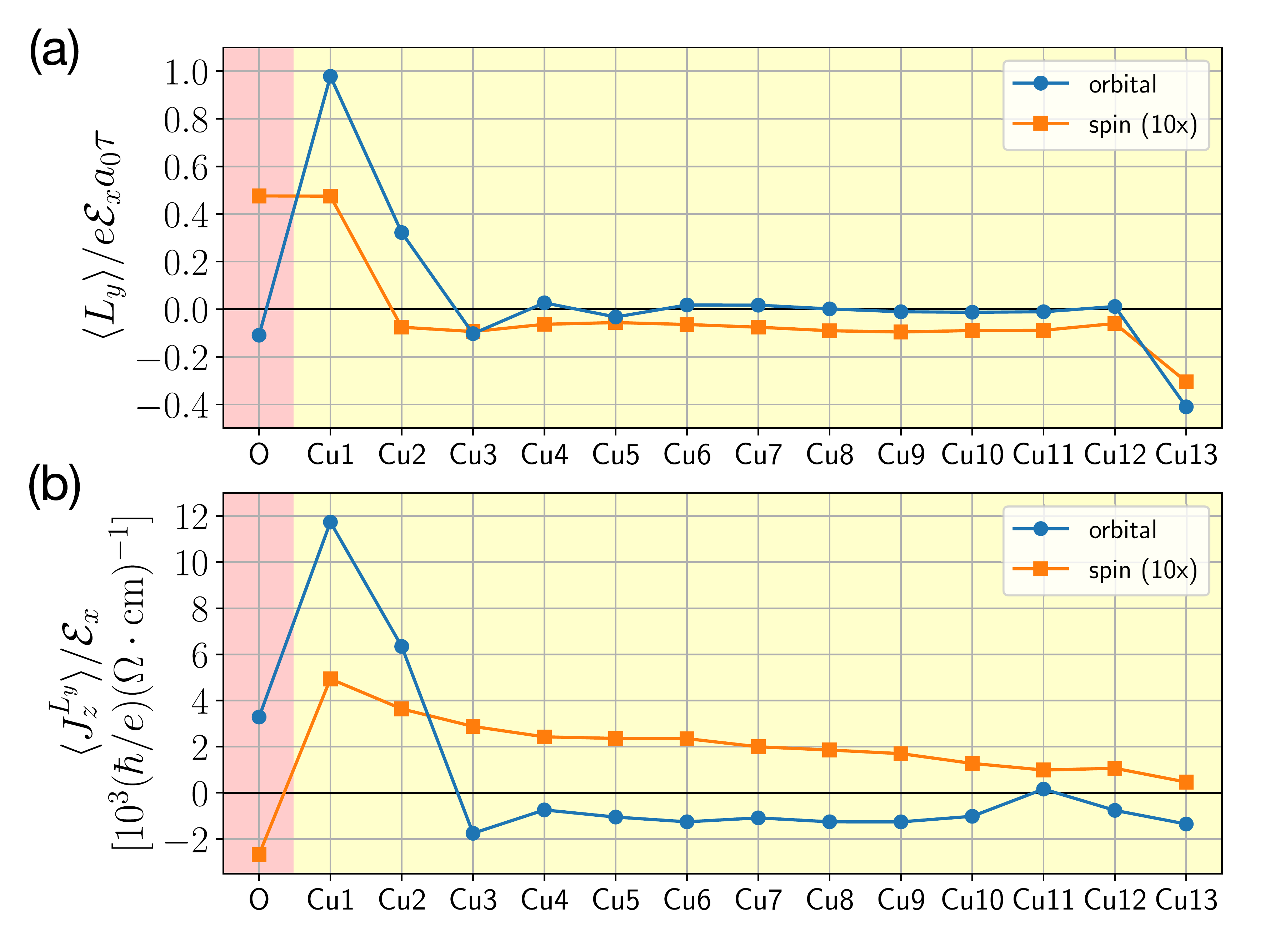}
\caption{
\label{fig:electric_response}
Electric field response of (a) OAM accumulation $\left\langle L_y \right\rangle$ and (b) orbital Hall current $\langle J_z^{L_y} \rangle$ (blue circles). Spin responses are also shown for comparison (orange squares), which are much smaller than the orbital counterparts. An external electric field $\mathcal{E}_x$ is applied along the $x$ direction and the $z$ direction defined such that it points toward the top surface (O/Cu1). The orbital responses are gigantic, comparable or larger than the spin responses in heavy elements.}
\end{figure}

Finally we investigate the consequences of an external electric field on OAM accumulation and orbital Hall current. The OAM accumulation is induced by the shift of the Fermi surface, which is known as orbital Rashba-Edelstein effect \cite{Go2017, Yoda2018, Salemi2019}. On the other hand, the orbital Hall current is generated by the intrinsic process of interband mixing \cite{Go2018}. For the calculation, we used the same method as in Ref. \cite{Go2020}. Figure~\ref{fig:electric_response}(a) shows the spatial profile of the $y$ component of the OAM accumulation $\left\langle L_y \right\rangle$ when the external electric field ($\mathcal{E}_x$) is applied along the $x$ direction. 
To make it dimensionless, we divide $\left\langle L_y \right\rangle$ by $e\mathcal{E}_xa_0 \tau$, where $a_0$ is the Bohr radius and $\tau$ is the momentum relaxation time. Positive sign of $\left\langle L_y \right\rangle$ is consistent with the clockwise orbital texture at the Fermi surface considering shift of the Fermi surface along the $-x$ direction. 
The magnitude of the orbital accumulation is gigantic as compared to spin accumulation at heavy metal surfaces. For example, the spin accumulation at W(110) surface 
is smaller by an order of magnitude~\cite{Go2020b}. Similarly, the orbital Hall current response $\langle J^{L_y} \rangle$, shown in Fig.~\ref{fig:electric_response}(b), also exhibits  a gigantic value, exceeding $\sim 10^4(\hbar/e)(\Omega\cdot\mathrm{cm})^{-1}$. This can be compared to the bulk spin Hall conductivity of Pt of $\sim 2.0\times 10^3(\hbar/e)(\Omega\cdot\mathrm{cm})^{-1}$ \cite{Guo2008}. In Figs.~\ref{fig:electric_response}(a) and \ref{fig:electric_response}(b), we also show the response of the spin accumulation and spin Hall current, respectively, for comparison. 
From this plot we observe that the current-driven orbital response is by roughly a factor of 20 more prominent than the spin response.

Although both OAM accumulation and orbital Hall current responses are pronounced near Cu1 and Cu2, they quickly decay as we move away from these surface Cu layers. It means that for the OAM to be injected into an adjacent ferromagnet, so that the orbital torque is exerted on the local moment, the Cu layer has to be ultrathin. This seem contradictory with the experimental observation of the unconventional SOT at the surface of oxidized Cu/ferromagnet structures, where the Cu layer is at least $10\ \mathrm{nm}$ thick \cite{Kim2020, Tazaki2020}. However, the samples of surface oxidized Cu used in experiments are far from being single crystalline and the oxidization level shows strong inhomogeneity \cite{An2016, Kim2020}. We speculate that part of the Cu film where the oxygen penetration is deep enough may exhibit strong ORE \emph{locally} near the ferromagnetic layer. Variation of the oxygen concentration is expected to affect spatial distribution of the oxygen, which alters effective area responsible for the SOT generation. Unfortunately, first principles investigations along this line go well beyond the scope of this work. Experimental effort to reduce the surface roughness of the Cu film may facilitate a direct comparison with our theoretical calculations.

In summary, we have shown that surface oxidization of Cu film leads to gigantic enhancement of the ORE for states near the Fermi surface as a result of the hybridization between the O $p$ orbital and the Cu $d$ orbital. The resulting chiral OAM texture in $\mathbf{k}$-space is found to be exceptionally large $\sim 0.5\hbar$ in its magnitude, which shows clockwise chirality. This leads to a large electric response of both OAM accumulation and orbital Hall current, and implies that the orbital torque mechanism may be significant at surfaces of oxidized Cu/ferromagnet structures. It also encourages experimental verification of the orbital texture in surface oxidized Cu through optical measurements such as angle-resolved photoemission spectroscopy.

\begin{acknowledgements}
D.G and Y.M. appreciate fruitful discussion with Shilei Ding, Mathias Kl\"aui, and Frank Freimuth. H.-W.L. thanks Chongze Wang, Jun‐Hyung Cho, and Myung Ho Kang for providing comments. We gratefully acknowledge the J\"ulich Supercomputing Centre for providing computational resources under project jiff40.
This work was funded by the Deutsche Forschungsgemeinschaft (DFG, German Research Foundation) $-$ TRR 173 $-$ 268565370 (project A11), TRR 288 $-$ 422213477 (project B06).
D.J. and H.-W.L were supported by Samsung Science \& Technology Foundation (Grant No. BA-1501-51).
\end{acknowledgements}


\bibliography{bib_CuO}

\end{document}